\documentclass[twocolumn,amsmath,amssymb,8pt]{revtex4}

\usepackage{graphicx}
\usepackage{dcolumn}
\usepackage{bm}
\usepackage[dvips]{epsfig}

\newcommand {\ket}[1] { |  #1 \rangle   }
\newcommand {\bra}[1] {  \langle #1  | }

\begin{document}

\title{Mixed Ab-initio Quantum Mechanical and Monte Carlo Calculations \\
of Secondary Emission from Si0$_2$ Nanoclusters.}

\author{Simone Taioli$^{*a,b}$, Stefano Simonucci$^{c}$, Lucia Calliari$^{a}$, Massimiliano Filippi$^{a}$, Maurizio Dapor$^{a,b}$}
\affiliation{$^{a}$\footnotesize{FBK-IRST Center for Materials and Microsystems, Via Sommarive 18, 38050
Povo (Trento), Italy \\
$^{b}$ The European Centre for Theoretical Studies in Nuclear Physics and Related Areas (ECT* Associates), Strada delle Tabarelle 286, I-38050 Villazzano (Trento) Italy\\
$^{c}$Department of Physics, University of Camerino, via Madonna delle Carceri 9, 62032 Camerino, Italy}}

\date{\today}

%%%%%%%%%%%%%%%%%%%%%%%%%%%%%%%%%%%%%%%%%%%%%%%%%%%%%%%%%%%%%%%%%%%%%

\begin{abstract}
A mixed Quantum Mechanical/Monte Carlo method (QMMC) for calculating Auger spectra from nanoclusters is presented. 
The approach, based on a cluster method, consists of two steps. 
Ab-initio quantum mechanical calculations are first performed to obtain accurate energy 
and probability distributions of the generated Auger electrons. 
In a second step, using the calculated line-shape as electron source, the Monte Carlo method 
is used to simulate the effect of inelastic losses on the original Auger line-shape. 
The resulting spectrum can be directly compared to `as-acquired' experimental spectra, 
thus avoiding background subtraction or deconvolution procedures. 
As a case-study, the O K-LL spectrum from solid SiO$_2$ is considered. 
Spectra computed before or after the electron has travelled through the solid, i.e. unaffected or affected by extrinsic energy losses, are compared to the pertinent 
experimental spectra measured within our group. 
Both transition energies and relative intensities are well reproduced. 
To the best of our knowledge, this is the first time that Auger spectra are computed ab-initio, with no free parameters involved. 
\end{abstract}

%%%%%%%%%%%%%%%%%%%%%%%%%%%%%%%%%%%%%%%%%%%%%%%%%%%%%%%%%%%%%%%%%%%%%
\maketitle

\section{INTRODUCTION}
In recent years, remarkable developments in two branches of
semiconductor physics have been achieved. One is the theoretical advance
in the study of band gaps and quasi-particle spectra in systems, notably transition-metal oxides,  
where electronic correlation creates states other than simple Fermi gas \cite{Hedin}. 
The other is the science of nanomaterials, whose electronic, spectroscopic
and transport properties bridge the gap between molecules and condensed matter \cite{Gerber}. 
The search for new technologies in nanoelectronics based on silicon and carbon
nanomaterials \cite{Kim,Jorio} has been fueled by the potential applications of such nanoclusters
as junctions in optical telecommunication networks or in 
integrated circuits where serious obstacles are represented by both the signal speed and spatial density.  
Despite many recent advances in the development of carbon based electronics \cite{Avouris},
silicon still represents the most widely used physical means. Furthermore, a miniaturization 
of silicon based devices would retain all our understanding of the physics 
of this semiconductor and the technological background of the microelectronics industry.  
But while the knowledge of the electronic properties of silicon and silicon oxides has advanced greatly,
largely due to the development of theories capable of treating the correlation energy \cite{Gunn},
theoretical results on the secondary emission and energy loss mechanisms are still the subject of discussion \cite{Werner}.

In this paper, we focus on the O K-LL spectrum from SiO$_2$ clusters. 
We present a combination of an ab-initio Quantum Mechanical method with Monte Carlo 
simulations (QMMC) to calculate non-radiative decay spectra from 
SiO$_2$ nanoclusters of different size taking into account post collisional interactions (PCI) due to electron dynamical screening. 
The method is validated by comparison to measured spectra obtained in our group. 
In this regard, we move along two independent directions. On the one hand, we start from the measured spectrum, affected by all kinds of energy losses suffered by the electron on its way out of the solid, and we recover the `true' Auger spectrum by conventional deconvolution procedures. On the other hand, using the computed O K-LL spectrum as electron source, we use the Monte Carlo method to simulate the effect of inelastic scattering on the original electron energy distribution. Key ingredient for both procedures is the inelastic scattering cross section or, equivalently, the Reflection Electron Energy Loss Spectrum (REELS).
While the well studied \cite{Ramaker2,Ramaker1,Riessen} O K-LL spectrum from SiO$_2$ clusters is taken here as a case study, 
we believe that our approach is sufficiently general to be applied to the interpretation of core electron spectroscopies, such as Photoemission (PES), 
Auger (AES) and Autoionization by taking into account dynamical screening effects in a variety of nanoclusters and condensed matter systems.

In order to gain new insights into the nature of correlation and of continuum states
in nanoclusters, a number of well-defined questions have to be addressed.
What is the nature of the fundamental process occurring in a scattering experiment, where an electron or a photon
is injected in the topmost layer?
What is the nature of the core-hole and double-hole states when electronic relaxations and charge disorder are included?
How does hole localization distort the double-hole density of states
(DOS) and manifest itself in the scattering process?
What is the role of dynamical screening in the Auger process?
Can the above considerations be combined to understand experimental data in a unified framework?
Are results size dependent?

Many models, described in several reviews \cite{Weissmann,Weightman1,Kleiman2,Ramaker3}, 
have been suggested to tackle the calculation of Auger spectra in solids, a difficult task due to the
high number of energy levels involved and the contribution of additional degrees of freedom which broaden
lineshapes. 
In particular, computational and theoretical procedures are represented            
by methods developed by Cederbaum \cite{Ceder,Ceder4}, Ohno \cite{Ohno} and Cini-Verdozzi \cite{Verdozzi}.
In each of these cases, Auger decay probabilities were computed by means of atomic Auger matrix elements
with the many-body Green function projected onto atomic states and the spectral lineshape obtained by calculating the residua
of such a projected quantity. This approach revealed very valuable in the calculation of
atomic and molecular \cite{Ceder2,Ceder3,Ohno1} Auger spectra but the extension to solids is fairly difficult and poor in results.
Verdozzi and Cini \cite{Verdozzi,Verdozzi1} improved such an approach to separate the atomic features
from those characteristic of a periodic structure (see below).
The striking difference between our approach and those used by all the above cited authors (and
further extensions of their work) lies in the computation of the many-body Green function.
In fact, in our method the many-body Green function is projected
onto `localized' states, which are a `local mixture' of atomic states.
Therefore, differently from previous works, our local projectors
are multicentered and may include some atomic space points.
In other words, the role played by our projectors is somehow similar
to that played by localized Wannier functions in solid state calculations.
It is clear that
the use of `localized' (not atomic) Auger matrix elements will prove to be very
valuable in the treatment of transitions, such as valence-valence-valence (V-VV),
where initial states are not atomic, but delocalized in nature.
The quest of multicenter projectors in the calculation of Auger matrix elements
is of paramount importance, because the use of atomic Auger matrix elements, acceptable
in molecular spectra, would result in a crude approximation for condensed matter core-hole spectroscopy.

As for core-valence-valence (C-VV) spectra, their understanding is presently based on the use of atomic 
transition matrix elements and it is rooted in Lander's original idea  \cite{Lander} that  such spectra should basically 
reflect the self-convolution of the valence band DOS, so called `band-like' spectra. 
While this is true for solids such as Li and Si, it fails for transition metals, Zn for example, 
whose C-VV spectra are instead referred to as `atomic-like'. 
Recognizing that correlation effects between the two final-state holes were responsible for such spectra, 
led to the Cini-Sawatzky theory \cite{Cini3,Sawatzky} which, 
derived at first only for systems with initially filled bands, 
has since become the framework for any quantitative interpretation of C-VV spectra. 
Within this model, hole-hole correlation energy is a semi-empirical parameter to be determined by comparison with measured spectra.
In particular, the effect of hole-hole repulsion on the O K-LL spectrum from bulk SiO$_2$ has been recently investigated 
by Van Riessen at al. \cite{Riessen}  using the Cini-Sawatzky model extended by Ramaker \cite{Ramaker1}. 
The spectrum is computed by multiplying experimental intra-atomic transition 
matrix elements times the distorted DOS due to hole localization.  
Despite the O K-LL spectrum could be satisfactorily reproduced, 
matrix elements and characteristic parameters were taken from atoms, 
a fact that can be inappropriate for ionic crystals, such as SiO$_2$, 
where occupation numbers are different from the atomic ones.

To the best of our knowledge, the calculation presented here represents the first attempt 
to address the questions listed above from first principles, with a
full ab-initio treatment of hole-hole correlation and taking into account collective excitations 
without free parameters, except for a constant factor to normalize spectral intensity. 
We will focus on the so-called principal or normal Auger process, whereas the contribution of satellites (associated with
shake processes) to the Auger spectrum is neglected.
\section{THEORY}
The general theoretical framework for interpreting Auger decay in nanoclusters 
is the theory of scattering \cite{Aberg}, where the initial state consists of a projectile, in our case 
typically photons or electrons, and a target in its ground state, while the final states are characterized by an ionized
system and one or more electrons asymptotically non interacting. 
The main problem to be addressed for the calculation of the Auger 
theoretical spectrum is the construction of a continuum electronic wavefunction that
accurately takes into account correlation effects among electrons bound in the nanoclusters and 
between the escaping and bound electrons.
To build this wavefunction we need to extend a method \cite{sim1,sim2} already successfully applied 
to the calculation of inner shell spectra in small molecules, 
such as CO \cite{taioli3} and C$_2$H$_2$ \cite{taioli2,taioli1}. 
However, the existence of several interacting decay paths,
quantum effects due to complex many-body interactions and localization/delocalization effects,
make such extension very difficult if one wants to keep the computational cost comparable to that of molecules.  

To compute the scattering wavefunctions 
$|\Psi_{\alpha,\varepsilon_{\alpha}}^->$, including the appropriate {\it incoming wave} 
boundary conditions, we use an approach based 
on Fano's multichannel theory \cite{Fano}, where the solution to the scattering 
process is represented by a linear combination of a metastable state $|\Phi>$ 
with states embedded in the non-interacting continuum 
many-fold $\{\chi_{\beta,\varepsilon_{\beta}}^-\}$, 
\begin{eqnarray}
\ket{\Psi^-_{\alpha,{\cal E}}}=\ket{\chi^-_{\alpha,\varepsilon_{\alpha}}}+  \nonumber \\ 
{M^-_\alpha(\varepsilon_{\alpha},E)\over E-E_r-i{\Gamma\over 2}}
\left[\ket{\Phi}+\lim_{\nu\rightarrow 0}\sum_{\beta}
\int_0^\infty{\ket{\chi^-_{\beta,\tau}}M^-_\beta(\tau,E)^*\over
E-E_\beta-\tau-i\nu}d\tau\right]  \nonumber \\
\label{TF9}
\end{eqnarray}
with $\Gamma$ and $M_\beta^-(\varepsilon_{\beta},E)$ defined as follows
\begin{eqnarray}
\Gamma&=&\sum_{\beta}\Gamma_\beta=2\pi\sum_{\beta}
|M^-_\beta(\varepsilon_{\beta},E)|^2
\label{TF10} \\
M_\beta^-(\varepsilon_{\beta},E)&=&{\bra \Phi}H-E\ket{\chi_{\beta,\varepsilon}^-};~\varepsilon_{\beta}=E_r-E_\beta
\label{TF11}
\end{eqnarray}
In equations (\ref{TF9}-\ref{TF11}) $E_r$ is the energy of the resonant state modified by 
the interaction with the continuum states $\{\chi_{\beta,\varepsilon_{\beta}}^-\}$ and $E_{\beta}$ 
is the energy of final states. 
These multichannel continuum wavefunctions can be obtained from a set 
of interacting continuum states $\{\chi_{\beta,\varepsilon_{\beta}}\}$, 
coupled by the interchannel interaction, by projecting
the Hamiltonian $H$ on a finite set of multicentered $L^2$ functions 
spanning the Hilbert space in the range of the ejected electron kinetic energy:
\begin{equation}\label{E999}
<\chi_{\beta\varepsilon_{\beta}}|\hat H-E|\chi_{\alpha\varepsilon_{\alpha}}>~
=~<\psi_{\varepsilon_\beta};~\beta|~\hat {\cal H}(E)~-~E~|\psi_{\varepsilon_{\alpha}};~\alpha>
\end{equation}
where $|\psi_{\varepsilon_\beta};~\beta>$ represents the antisymmetric tensorial product of the escaping electron wavefunction
at energy $\varepsilon_{\beta}$ with the doubly ionized target wavefunction ($\beta$).
In equation (\ref{E999}) $\hat {\cal H}$ is the projected Hamiltonian, which includes 
the projected scattering potential
\begin{equation}\label{E1000} 
\hat V^{\beta,\alpha}(\varepsilon_{\beta},\varepsilon_{\alpha})=<\chi_{\beta\varepsilon_{\beta}}|~\hat V |\chi_{\alpha\varepsilon_{\alpha}}> 
\end{equation}
describing the interchannel interaction among bound and emitted electrons in the scattering region.   

To find an accurate solution of equation (\ref{E999}) in SiO$_2$ nanoclusters, a number of 
numerical and theoretical techniques have been implemented. 
The first problem is the choice of the basis set for the bound states and the scattering 
wavefunction.
Due to the large number of atoms and open channels in comparison to molecules, 
the size of the Hilbert space where the wavefunctions are expanded increases rapidly for nanoclusters. 
An increasing number of Gaussians may lead to convergence    
to non-physical states due to basis set superposition errors and linear dependence problems \cite{Spack}, particularly in 
the calculation of metastable states. 
In order to keep the number of excited determinants in the doubly ionized
target low, we split the problem in two steps:     
we first calculate the bound orbitals in a restricted Gaussian basis set; 
we then introduce an auxiliary basis
set of modified Gaussian functions to compute the matrix elements 
of the scattering potential (equation (\ref{E1000})).
The Hartree-Fock procedure takes the variationally optimized atomic wavefunctions of 
different components of the nanocluster as a starting point for the calculation of the nanocluster
wavefunctions, ensuring the convergence of the multiconfiguration procedure \cite{Szabo}
to the appropriate ionic states within the space spanned by a minimal basis set. 
An adequate representation of the continuum orbital, comparable to the solution of the 
multiconfiguration procedure in bound orbitals, forces us to enlarge the basis set in scattering
wavefunction calculations. 
We emphasize that the scattering basis set is built with a diffuse Gaussian basis set multiplied 
by symmetry adapted Hermite polynomials to reproduce the oscillating behaviour
of the continuum wavefunctions inside the scattering region. 
This enlarged basis set is used both in the expansion of the continuum orbital and in 
the projection of the interchannel coupling inside the Hilbert space spanned 
by the interacting decay channels.
Continuum orbitals have to be orthogonal to bound orbitals.
To satisfy this constraint one can use two different methodologies.
One is direct orthogonalization, the other uses projection of the kinetic operator onto the 
orthogonal complement of the bound states. The latter procedure defines an effective
potential, which has the same structure of the Phillips-Kleinman potential \cite{PhiKlei} used 
in condensed matter calculations for building valence orbitals orthogonal to the core states.  
Surprisingly, we found that the Phillips-Kleinman technique 
leads to numerical instabilities, with a basis set dependence, in the calculation of the SiO$_2$ nanoclusters 
continuum density of states. 
It turned out that such instability is found whenever the eigenvalues of the
metrics of the chosen basis set are smaller than $10^{-4}$.
Attention has to be paid, when choosing the basis set, to describe appropriately
the scattering potential in the energy region where electrons are emitted and to
avoid a small overlap between the basis elements.  
After taking into account such a criterion, 
calculations of the continuum density of states for SiO$_2$ nanoclusters 
gave the same results in terms of either approach used to satisfy the orthogonality constraint. 

Due to the enormous number of decay channels coupled with the final        
states, the Auger probability in nanocluster calculations
is spread all over the bands with a complex distribution. 
Accordingly, the computational cost to evaluate the interchannel potential in equation (\ref{E1000}), 
responsible for the interaction between the open channels and for the probability spread
among the non interacting channels, may be so high that it could not be directly 
performed as previously done for molecules \cite{sim3}.
Unfortunately, interchannel interaction cannot be neglected, because its absence 
is the main source of errors and of large
discrepancies with experimental spectra in scattering calculations \cite{sim3}. 

One can argue that the strong ionic bond between Si and O leads to a rather large
interaction. The electronic cloud distortion, with silicon to oxygen charge donation,
favors holes trapping on the oxygen site.
Moreover, small multisite interaction among the SiO$_2$ units, 
is responsible for the narrow valence bandwidth.
As a consequence, one expects a very low hole mobility, clamped down by a large hole-hole interaction. 
From these observations, one expects that the final states are created by intra-atomic transitions with 
the two holes strongly localized.

In light of what we have shown above, a theoretical approach for calculating atomic-like and band-like
Auger lineshapes in nanoclusters with strong correlation effects is proposed here.
In this method both the initial state core-hole and the final state valence holes are first 
localized on the central atom and multisite interactions
enter naturally in the model Hamiltonian as a perturbation.
Therefore, a number of localized states, which account for the majority of the decay probability,
is selected among the open channels. These `tight binding' states are further coupled via the interchannel
potential in such a way that the model Hamiltonian can be diagonalized at a very high level of accuracy 
in a smaller space, corresponding to the most intense open channels.
This calculation results in a number of non interacting states,
which account for the quasi-atomic behaviour of the nanocluster.
The long range or band-like behaviour of the Auger decay is calculated at this stage 
by switching on the weak multisite interaction.
The perturbative coupling of each non-interacting channel 
with a number of states describing transitions close in energy, 
allows us to choose, among the degenerate states, those presenting a maximum overlap with the 
selected channel and, therefore, to obtain the whole lineshape.  
\section{CASE STUDY: SiO$_2$}
\begin{figure}[hbtp]
\centerline{\includegraphics[width=9cm]{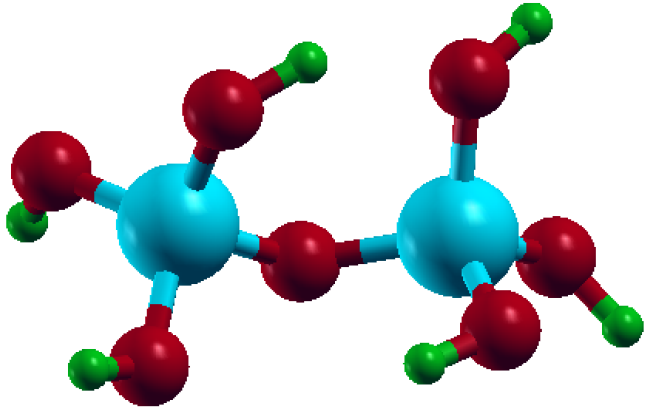}} 
\caption[]{`(Color online)' Sketch of the SiO$_2$ nanocluster optimized structure. Oxygens are in red, silicons in light blue,
hydrogens in green.}
\label{geometry}
\end{figure}

To test our approach, we take the Si$_2$O$_7$H$_6$ nanocluster as representative of bulk SiO$_2$ 
to calculate the normal O~K-LL Auger spectrum. 
Within the Born-Oppenheimer approximation, the optimized structure of Si$_2$O$_7$H$_6$
with a central oxygen atom has been obtained by minimizing the free energy of the electrons at each nuclei position
with a smooth Fermi-Dirac smearing, relaxing the atomic bond length
until the forces are less then 0.01 eV/\AA \quad per atom. 
The resulting atomic configuration is plotted in Fig. (\ref{geometry}).
Stars of neighbors, centered on the central oxygen atom,
are found at a distance of $3.103 ~a_0$ for silicons and $4.953~a_0$ for oxygens.  
The present results are in good agreement with previous 
calculations on band structure and geometry of similar Si0$_2$ clusters \cite{Yip}. 
Following our scheme, correlation in the initial state of the ionized target, characterized by
a single inner vacancy in the $1s$ orbital of the central oxygen, and in the final states
of the doubly ionized nanocluster, characterized by two vacancies in the valence orbitals,
has been treated by extended configuration interactions (CI) expansion using 
the $3s,3p,3d$ orbitals of silicon and
$3s,3p,3d$ of oxygen.
Calculations of the bound states have been carried out using a basis set of Hermite Gaussian functions with (s, p, d)-type
character centered on the nuclei. The basis set has been obtained by a variational procedure, optimizing the exponents
of the standard 6-31G$^*$ basis set. In more detail we scan a finite range of values above and below the proposed 6-31G$^*$ ones, taking the
values which lower the total energy.
An auxiliary basis set, more dense and enlarged with diffuse functions, has been used
to represent properly the projected potential $V^{\beta, \alpha}$ in equation (5), until variational stability
of the Auger decay rate with respect to changes in the basis set is reached. This criterion
allows an accurate description of the continuum orbitals around the Auger decay energies.
The scattering calculations highlight the fundamental role of d-symmetry functions in the
expansion of the projected potential.
The resulting basis set is made up of $[9s (= 4s_{Si}+3s_{O}+2s_{H}) +  6p (= 3p_{Si}+2p_{O}+1p_{H}) + 3d (= 1d_{Si}+2d_{O})]$
contracted Hermite Gaussian functions, for a total of 215 contracted Gaussians.
The exponents $\alpha_s, \alpha_p, \alpha_d$ are in the range $[\alpha_s^{Si}=0.001 \div 1.0, \alpha_p^{Si}=0.004 \div 1.2, \alpha_d^{Si}=1.0]$
for the (s, p, d)-type Gaussians on silicons, $[\alpha_s^{O}=0.002 \div 0.5, \alpha_p^{O}=0.07 \div 1.0, \alpha_d^{O}=1.0]$
for the (s, p, d)-type Gaussians on oxygens, $[\alpha_s^{H}=0.03 \div 1.0, \alpha_p^{H}=1.0]$ for the (s, p)-type Gaussians on hydrogens.
We adopted a larger basis set on the central oxygen with $[4s + 4p + 3d]$ to properly represent hole localization effects.
The examined nanocluster has a ionic character, with the central oxygen atom in the Si$_2$0 
unit found in the Si$^{+1.49}$O$^{-1.20}$ charge configuration.
We found that the Si$_2$O$_7$H$_6$ nanocluster correctly reproduces one electron 
properties, localization effects and the charge population of solid SiO$_2$. 
In table (\ref{table1}) we have reported the energies, referred to the vacuum level, and the relative decay rates of the most intense
fifteen transitions, all with double holes localized on the central oxygen atom, obtained after switching on the interchannel coupling.
\begin{table}
\caption{\label{table1} Kinetic energies ($E_{kin}$) in eV, referred to the vacuum level, 
and probabilities ($\Gamma_{\alpha}$) in arbitrary units
for the O K-LL Auger localized states of Si$_2$O$_7$H$_6$ nanocluster
according to the double holes configu-rations (O K-LL) in the  central oxygen atom 
and final total spin $S^{2}$ 
(0=singlet, 1=triplet).}
\begin{center}
\begin{tabular}{c|c|cc|c|c|cc}
O K-LL & $S^{2}$ & $E_{kin}$ & $\Gamma_{\alpha}$ & O K-LL & $S^{2}$ & $E_{kin}$ & $\Gamma_{\alpha}$ \\
\hline
$2s-2s$ & (0) & 458.75 & 0.281 & $2p-2p$  & (0) & 497.89 & 0.395 \\ 
$2s-2p$ & (0) & 473.4 & 0.252 & $2p-2p$ & (0) & 498.74 & 0.425 \\ 
$2s-2p$  & (0) & 477.41 & 0.322 & $2p-2p$ & (1) &  499.88 & 0.003 \\
$2s-2p$  & (0) & 477.99 & 0.308 & $2p-2p$  & (0) & 500.28 & 0.409 \\
$2s-2p$ & (1) & 481.64 & 0.077 & $2p-2p$  & (0) & 501.98 & 0.481  \\
$2s-2p$  & (1) & 484.96 & 0.090 & $2p-2p$  & (0) & 502.45 & 0.493 \\
$2s-2p$  & (1) & 485.73 & 0.094 & $2p-2p$  & (1) & 504.13 & 0.003 \\
$2p-2p$ & (0) & 493.94 & 0.330 & & & & \\
\end{tabular}
\end{center}
\end{table}
\section{EXPERIMENTAL DATA HANDLING}
Since first-principles methods necessarily involve
approximations, the most important of which, in scattering calculations, is the approximate form of the 
electron wavefunction in the continuum, it is important to validate the calculations
against experimental data. 
To this end, O K-LL spectra were measured from a 36 nm thick SiO$_2$ film, grown on a Si substrate. 
The film surface was cleaned by annealing at 800$^{\circ}$C in a Ultra High Vacuum 
preparation chamber connected to the analysis chamber (base pressure 2$\times 10^{-10}$ mba). 
The analysis chamber is a PHI545 instrument equipped with double-pass Cylindrical Mirror Analyzer, 
coaxial electron gun, non-monochromatic Mg K$\alpha$ X-ray source and He discharge lamp.
The X-Ray (h$\nu$=1253.6 eV) excited O K-LL spectrum was recorded at constant 
analyzer energy resolution, $\Delta$E=0.6 eV, as measured on the Pd Fermi edge 
of a HeI (h$\nu$=21.2 eV) excited valence band photo emission spectrum. 
A Reflection Electron Energy Loss Spectrum, excited from the same sample 
by 500 eV electrons, i.e. near the kinetic energy of O K-LL electrons, was acquired 
within the same instrument and at the same analyzer resolution of the Auger spectrum. 
To keep electron beam induced damage to a minimum, the current density was lower 
than 40 A/m$^2$ and the acquisition time was lower than 6 minutes. 
The energy scale, was calibrated on clean Au and Cu, following recommended procedures \cite{Seah1999}. 
Kinetic energies are referred to the vacuum level.

\begin{figure}[hbtp]
\centerline{\includegraphics[width=10cm]{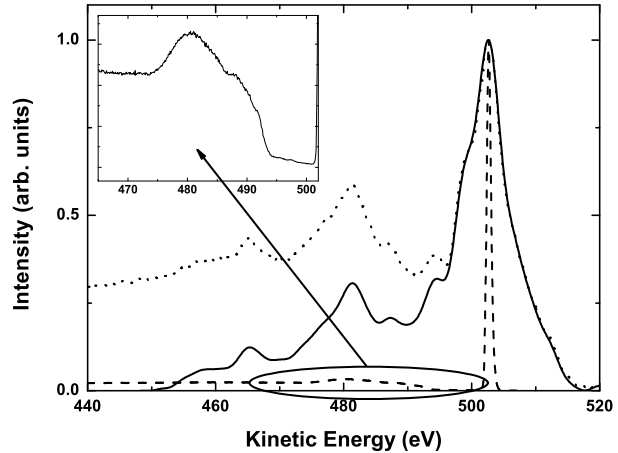}}
\caption[]{Plot of the experimental data. 
The X-Ray (h$\nu$=1253.6 eV) excited O K-LL spectrum (continuous line) has been obtained by
deconvolution of a REEL spectrum (dashed line) from a spectrum obtained subtracting a
constant background, defined by the region immediately to the right (520-530 eV) of the main peak
(dotted line).}
\label{experimental}
\end{figure}

`As acquired' spectra were corrected for the energy dependence of the analyzer transmission function. 
The latter had been previously determined on clean Au, Ag and Cu as explained in \cite{Seah1995}.
A constant background, defined by the region immediately to the right (520-530 eV) of the main peak in the O K-LL spectrum, 
was subtracted, and intensity was normalized to unit height of the main peak, 
resulting in the spectrum (dotted line) shown in Fig. (\ref{experimental}). Also plotted in Fig. (\ref{experimental}) is the REEL spectrum (dashed line) which includes the so 
called elastic or zero loss peak together with the associated inelastic losses. 
The spectrum is slightly shifted in energy, so that the zero loss peak is 
aligned to the main O K-LL peak. 
Its intensity is normalized to unit height of the zero loss peak. A vertical expansion of the loss features
is shown in the inset. 
By deconvoluting the REEL spectrum from the measured O K-LL spectrum, 
the O K-LL line-shape given by the continuous line in Fig. (\ref{experimental}) was finally obtained.

\section{LINESHAPE ANALYSIS}
\begin{figure}[hbtp]
\centerline{\includegraphics[width=10cm]{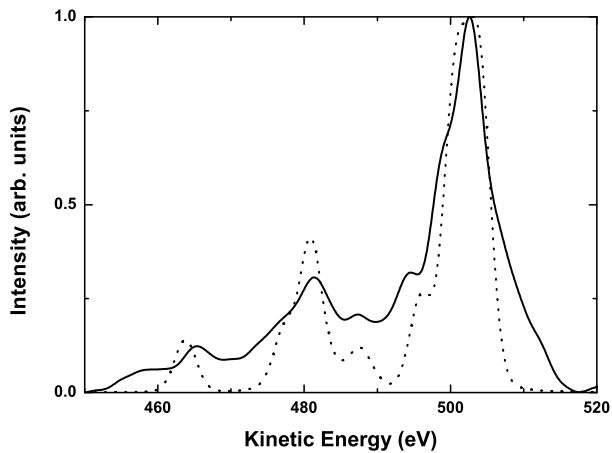}}
\caption[]{O K-LL Auger spectrum in Si$_2$O$_7$H$_6$. Comparison between the Quantum Mechanical (QM) 
theoretical data (dotted line) and the experimental data after deconvolution of a REEL spectrum
(continuous line).}
\label{comparison}
\end{figure}

This spectrum is compared to the ab-initio calculated O K-LL spectrum in Fig. (\ref{comparison}).
The computed
spectrum is generated by convoluting the theoretical transition rates with a Voigt profile to take into account both the
bandwidth of the X-ray source and the finite resolving power of the electron spectrometer. This results in a convolution of a Gaussian
with  $\Sigma_g$ = 1 eV, according to our instrument resolution, and a Lorentzian with $\Sigma_l$ = 0.1 eV, which is the computed
total Auger decay rate.
Both spectra are normalized to a common height of the main peak, while no energy shift was required to obtain the kind
of agreement shown in the figure.
We observe that our calculation correctly reproduces transition energies.
These are distributed over three different regions.
From right to left, features between $[494-510]$ eV are due to $2p-2p$ hole configurations in the final states (K-L$_{23}$L$_{23}$ transitions);
the most intense peaks are singlet transitions around $[501-503]$ eV (maximum at $502.76$ eV), close to the 
experimental peak ($502.58$ eV). 
Between $[475-492]$ eV a second group of features is found, where the hole configuration in the final states is
$2s-2p$ (K-L$_{1}$L$_{23}$ transitions);
the most intense peaks are singlet transitions around $[480-482]$ eV (maximum at $481.1$ eV), close to the
experimental peak ($481.38$ eV).
Finally a third group of features is observed between $[460-470]$ eV, where the hole configuration in the final states is 
$2s-2s$ (K-L$_{1}$L$_{1}$ transitions);
the most intense peaks are singlet transitions around $[463-465]$ eV (maximum at $463.86$ eV), lower than the
experimental peak ($465.25$ eV).
In general the most intense peaks are singlet transitions with both holes on the central oxygen
atom, whereas the shoulders are triplet transitions, unfavored by selection rules. 

Besides transition energies, also calculated values of the decay rates
are satisfactorily reproduced in Fig. (\ref{comparison}), meaning that the adopted scattering basis set is
appropriate to represent the projected potential within the scattering region for any channel.  
However, the ss and sp contributions are slightly overestimated.
This is due to lack of electronic correlation or to dynamic screening effects \cite{Ramaker3}
in the CI calculation.
The size of the excited determinants space in the CI procedure for both the
intermediate core-hole and double ionized final states
has been chosen large enough to reproduce well transition energies and decay probabilities, but
a larger active space should be, in principle, adopted
to take into account hole delocalization on all valence bands and electronic excitations.
Furthermore, the localized (not atomic) Auger matrix elements are kept fixed during
this procedure and their square modulus enters as a multiplicative
factor in the lineshape analysis. In principle,
they should be updated with a new electronic continuum wavefunction derived
by a multichannel procedure, taking into account the correlation between the double ion
and the new escaping electron wavefunction.
Larger CI calculations, though computationally very expensive, would correct the
lineshape for the overestimation of the s to p contribution.
The same kind of agreement seen in Fig. (\ref{comparison}) is obtained by comparing our calculated O K-LL Auger spectrum to the measured spectrum
by van Riessen et al. \cite{Riessen} and Ramaker et al. \cite{Ramaker2}.
Without loss of accuracy the present method thus allows us to calculate the 
Auger lineshape in nanostructures and solids at a computational cost comparable to that typical for atoms and molecules.
\section{MONTE CARLO SCHEME: QMMC VS. MEASURED SPECTRUM}
As stated above, the comparison between computed and experimental spectra requires 
to account for changes caused to the original electron energy distribution by energy 
losses suffered by the Auger electron on its way out of the solid. 
The typical procedure in this respect is the one described above. 
Alternatively however, one could simulate the effects of inelastic losses on the `original' distribution.
The basic idea is to use the ab-initio Auger 
probability distribution as a source of electrons which undergo
inelastic processes. 
Within our model the ab-initio spectrum could include the intrinsic energy loss, due to shake processes.
As stated above, however, shake processes were neglected in the present paper.
On the other hand, extrinsic energy losses,
due to the interaction with collective degrees of freedom of the valence charge (valence
plasmons \cite{Cini1}) and to multiple inelastic scattering, should be dealt with after computing the spectrum and using statistical methods.
In this way inelastic background would be superimposed to the theoretical spectrum instead of removing it, a procedure not free from
uncertainty.
Possible effects of background subtraction on the spectrum final shape are discussed, for example, in \cite{Zakhvatova}.

\begin{figure}[hbtp]
\centerline{\includegraphics[width=10cm]{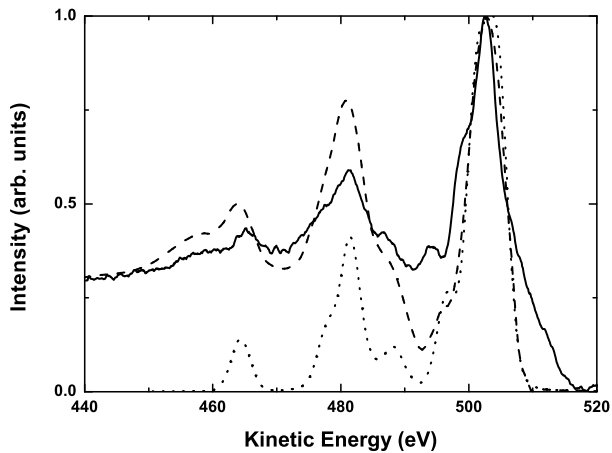}}
\caption[]{O K-LL Auger spectrum in Si$_2$O$_7$H$_6$. Comparison between the Quantum Mechanical theoretical data (dotted line), the Monte Carlo 
results (dashed line) and the original experimental data (continuous line).}
\label{comparison2}
\end{figure}

The simplest approach to account for the extrinsic loss is the Monte Carlo method \cite{Ding}.
Details of the Monte Carlo scheme describing electron energy losses in silicon dioxide 
were given in previous works \cite{dapor2003,Dapor2006,dapor2}. 
In the following we will briefly summarize the numerical approach used in this paper. 
The theoretical
Auger spectrum previously calculated is assumed to be the initial energy distribution of the escaping electrons.
The stochastic process for multiple scattering follows a Poisson-type law. 
The step-length $\Delta s$ is given by $\Delta s = -\lambda\;\ln(\mu_1)\;,$ 
where $\mu_1$ is a random number uniformly distributed in the range $[0,1]$. 
$\lambda$ is the electron mean free path:
\begin{equation}
\lambda(E)=\frac{1}{N[\sigma_{el}(E)+\sigma_{inel}(E)]}\;,
\end{equation}
where $N$ is the number of SiO$_2$ molecules per unit volume, $\sigma_{el}(E)$ is total elastic scattering cross section,
and $\sigma_{inel}(E)$ is the total inelastic scattering cross section at the impinging electron kinetic energy $E$. 
The calculation of differential and total elastic scattering cross sections
was performed by the Relativistic Partial Wave
Expansion method \cite{jablonski2004,dapor1996,dapor2003}, whereas
differential and total inelastic scattering cross sections
were computed using the Ritchie theory \cite{ritchie}. 
Taking Buechner experimental optical data \cite{buechner} 
for the evaluation of the long-wavelenght limit of the dielectric function
$\varepsilon(\omega)$, one obtains the differential inelastic scattering cross section as:
\begin{equation}
\frac{d\sigma_{inel}(E,\omega)}{d\omega}=\frac{me^{2}}{2\pi\hbar^{2} N E}\mbox{Im}\left[\frac{-1}{\varepsilon(\omega)}\right]
S\left(\frac{\omega}{E}\right)\;,
\end{equation}
where $\omega$ is the energy loss and, according to Ashley \cite{ashley88}, the function $S$ is given by 
\begin{equation}
\label{essefunction}
S(x)=(1-x)\ln\frac{4}{x}-\frac{7}{4}x+x^{3/2}-\frac{33}{32} x^{2}\;.
\end{equation}
Before each collision, a random number $\mu_2$ uniformly 
distributed in the range $[0,1]$ is generated and compared with the
probability of inelastic scattering $q_{inel}=\sigma_{inel}/(\sigma_{inel}+\sigma_{el})\;.$
When $\mu_2$ is smaller or equal to the probability of inelastic scattering, 
then the collision is inelastic; otherwise, it is elastic.
If the collision is elastic, $\theta$, the polar scattering angle, is selected 
so that the integrated scattering probability in the range $[0,\theta]$ is equal to 
the random number $\mu_3$ uniformly distributed in the range $[0,1]$:
\begin{equation}
\mu_3=\frac{1}{\sigma_{el}}\int_0^{\theta}\frac{d\sigma_{el}}{d\Omega}\;2\pi\;\sin\vartheta\;d\vartheta\;,
\end{equation}
where $\Omega$ is the solid angle of scattering.
If the collision is inelastic, the energy loss $W$ of an incident electron with kinetic energy $E$ 
is computed via a random number $\mu_4$, uniformly distributed in the range $[0,1]$:
\begin{equation}
\mu_4=\frac{1}{\sigma_{inel}}\int_0^W \frac{d\sigma_{inel}}{d\omega}d\omega\;,
\end{equation}
Auger electron generation is simulated assuming a constant depth distribution whose thickness, according to \cite{Riessen}, was set to 40 \AA.
To obtain good accuracy in the results the number of random walks is $10^8$.
A plot of the QMMC calculation compared to the experimental data before deconvolution of energy losses
is given in Fig. (\ref{comparison2}). 
The original theoretical spectrum is also shown for reference.
One can see that QMMC enhances and broadens the Auger probability 
increasingly upon decreasing the electron kinetic energy. 
The large broadening of the K-L$_1$L$_{2,3}$ peak after Monte Carlo treatment 
is due to the main plasmon of SiO$_2$ (see inset in Fig. (\ref{experimental})), 
whose distance from the zero loss peak is the same as the distance between the K-L$_{23}$L$_{23}$ and K-L$_{1}$L$_{23}$ 
features in the Auger spectrum. 
\section{CLUSTER SIZE EFFECTS}
Since we are dealing with nanoclusters, it makes sense to assess the dependence 
of correlation effects on the cluster size.
It is well known that an increasing number of atoms in the cluster causes the one-electron DOS to approach the
band DOS of the solid. 
\begin{figure}[hbtp]
\centerline{\includegraphics[width=10cm]{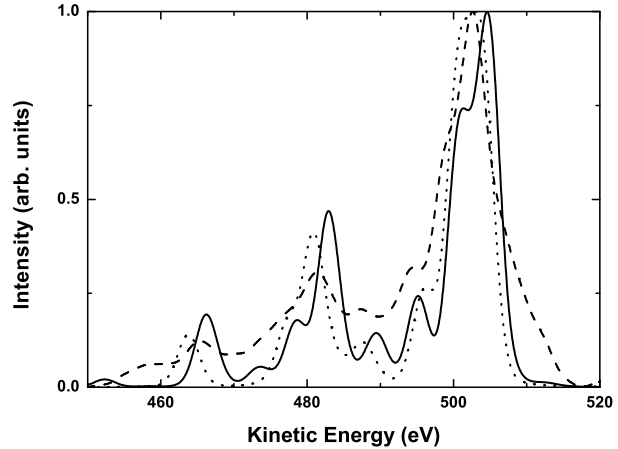}}
\caption[]{Comparison between the O K-LL spectrum in Si$_2$OH$_6$ (dotted line), Si$_2$O$_7$H$_6$ (continuous line) and
experimental data (dashed line).}
\label{comparison3}
\end{figure}
Increasing the number of atoms to simulate a bulk situation enlarges
the number of available transitions to the final states to a point where the computational cost is enormous, 
while no appreciable change in the spectroscopic observables is obtained.
In fact, the quasi-atomic nature of Auger emission found in our ab-initio calculations
forces the hole-hole DOS to be centered on the atom where the initial core-hole has been created.
We have investigated the role of cluster size by carrying out
calculations on a smaller cluster of atoms, Si$_2$OH$_6$,
to see how it compares with the bigger cluster.
This nanocluster represents the minimal cluster size which still describes the chemical environment
found in solid SiO$_2$.
In Fig. (\ref{comparison3}), we compare the calculated O K-LL spectrum of Si$_2$OH$_6$ (dotted line) to that of Si$_2$O$_7$H$_6$ (continuous line). 
The experimental spectrum is also superimposed (dashed line). All spectra are normalized to a common height of the main peak, while no energy shift is performed.  
The two computed spectra were obtained by convolution with the same Voigt profile. 
We see that the main features of the spectrum are
already consistently reproduced in the smaller cluster, though with a shift 
towards high energies. This `chemical shift'
decreases from 2.52 to 1.45 eV on moving from the low energy to the high energy peak.
This is a predictable
consequence of the fact that Si dangling bonds are saturated with different elements for the two clusters (H for the smaller
cluster and O for the bigger one)
with a non-local dependence which is larger for lower
angular momenta. 
Moreover, we observe that the Si$_2$O$_7$H$_6$ lineshape is less resolved than the corresponding Si$_2$OH$_6$ one, 
as expected due to the increased number of decay channels. 
Though the two computed spectra do not differ much in the number of features and in relative decay rates, 
the overall agreement with the experimental spectrum slightly favors the bigger cluster. 
This observation confirms that our cluster approach is well-suited for the calculation of Auger lineshapes from bulk materials and
that the 15 atom cluster considered in the present paper is near to the optimal size.
In conclusion, we believe that enlarging further the cluster would not change substantially
the Auger spectral lineshape as expected on the basis of the local nature of AES. Inclusion, in the computed spectrum, of 
shake-up and shake-off transitions and of
interactions between crystal phonons and escaping electron would result in a broadening of the lineshape, correcting
for the difference between the computed and experimental linewidth. 
\section{CONCLUSIONS}
We have demonstrated the potentiality of our first principle model 
for calculating and predicting the lineshapes 
of X-Ray excited Auger Electron Spectra from silicon dioxide nanoclusters. 
Furthermore, a Monte Carlo analysis has been applied to the Quantum Mechanical 
computation of spectra to mimic the electron energy loss in condensed matter. 
Comparison to our experimental data reveals a satisfactory agreement, 
in the energy position and relative intensities of peaks, and 
in the background contribution over the entire investigated energy range.

\end{document}